\pdfoutput=1
\documentclass[11pt]{article}

\usepackage[final]{acl}

\usepackage{times}
\usepackage{latexsym}
\usepackage{graphicx}  % For including images
\usepackage{subcaption} % For subfigures
\usepackage{makecell}

\usepackage{multirow}
\usepackage{booktabs}
\usepackage{hyperref}

\usepackage[T1]{fontenc}
\usepackage[utf8]{inputenc}

\usepackage{microtype}

\usepackage{inconsolata}

\usepackage{graphicx}

\usepackage{tcolorbox}
\usepackage{xcolor}

\definecolor{promptblue}{RGB}{210,230,250}
\definecolor{promptgreen}{RGB}{210,250,210}
\definecolor{promptyellow}{RGB}{250,250,210}
\definecolor{promptred}{RGB}{250,210,210}
\definecolor{promptpurple}{RGB}{230,210,250}
\definecolor{promptorange}{RGB}{255,228,196}
\definecolor{promptcyan}{RGB}{224,255,255}
\definecolor{promptgray}{RGB}{230,230,230}

\title{Challenges for AI in Multimodal STEM Assessments: a Human-AI Comparison}

\author{
  Aymeric de Chillaz\textsuperscript{1}\thanks{Both authors contributed equally to this research.} \quad
  Anna Sotnikova\textsuperscript{1}\footnotemark[1]\thanks{Corresponding author: \texttt{aasotniko@gmail.com}} \quad
  Patrick Jermann\textsuperscript{1} \quad
  Antoine Bosselut\textsuperscript{1} \\
  \textsuperscript{1}EPFL
}

\begin{document}
\maketitle
\begin{abstract}

Generative AI systems have rapidly advanced, with multimodal input capabilities enabling reasoning beyond text-based tasks. In education, these advancements could influence assessment design and question answering, presenting both opportunities and challenges. To investigate these effects, we introduce a high-quality dataset of 201 university-level STEM questions, manually annotated with features such as image type, role, problem complexity, and question format. Our study analyzes how these features affect generative AI performance compared to students. We evaluate four model families with five prompting strategies, comparing results to the average of 546 student responses per question. Although the best model correctly answers on average 58.5\% of the questions using majority vote aggregation, human participants consistently outperform AI on questions involving visual components. Interestingly, human performance remains stable across question features but varies by subject, whereas AI performance is susceptible to both subject matter and question features. Finally, we provide actionable insights for educators, demonstrating how question design can enhance academic integrity by leveraging features that challenge current AI systems without increasing the cognitive burden for students.
\end{abstract}

\section{Introduction}

Generative AI has been widely tested in educational applications, including its ability to answer exam-level questions~\cite{Malik2023, lan2024surveynaturallanguageprocessing, wang2024largelanguagemodelseducation}. There are two key challenges: AI can be misused in ways that undermine fair assessment, and its mistakes often appear convincing, potentially misleading students~\citep{borges2024, wang2023scibench, zhong2023agieval, arora2023llms}. To better understand these risks, benchmarks were introduced to assess AI performance~\citep{wang2024scibenchevaluatingcollegelevelscientific}.
%A key challenge is that AI can generate convincingly incorrect responses, requiring students to critically evaluate its outputs while also posing risks of misuse that could compromise fair assessment~\citep{borges2024, wang2023scibench, zhong2023agieval, arora2023llms}. As a step to address these concerns, benchmarks were introduced to evaluate AI performance~\citep{wang2024scibenchevaluatingcollegelevelscientific}.  
%Generative AI has been widely tested in various educational applications ~\cite{Malik2023, lan2024surveynaturallanguageprocessing, wang2024largelanguagemodelseducation}. One major research focus is on assessing AI’s ability to answer exam-level questions. A key challenge is that students must learn to identify convincingly incorrect responses, while educators face risks of AI misuse, which may compromise fair assessment ~\citep{borges2024, wang2023scibench, zhong2023agieval, arora2023llms}. As an action towards solving this issue, researchers started to introduce benchmarks to measure AI capabilities~\citep{wang2024scibenchevaluatingcollegelevelscientific}. This allows us to understand what are the potential risks for students and how teachers should adjust their materials.
%\citet{wang2024scibenchevaluatingcollegelevelscientific} introduced a college-level benchmark with both text-only and image-based questions. As AI capabilities continue to advance, it is crucial to evaluate performance on increasingly complex questions.
\begin{figure}[t]
  \includegraphics[width=\columnwidth]{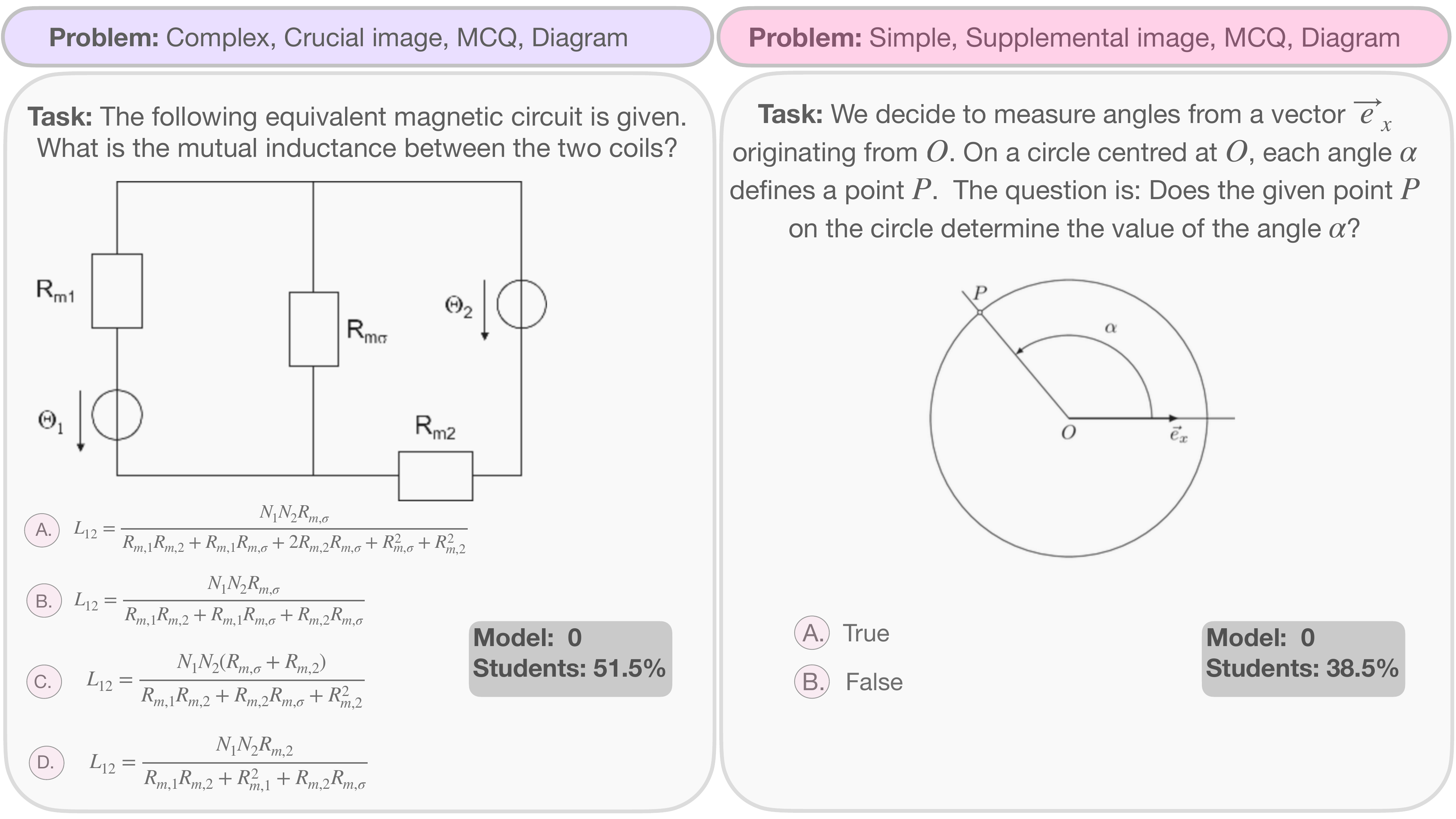}
  \caption{Example of STEM problems with average model performance (majority vote) compared to average student performance.}
  \label{fig:main}
\end{figure}
Recent advances in multimodal large language models (LLMs) have led to extensive efforts in developing image-based exam datasets, particularly in STEM. \citet{anand2024mmphyqamultimodalphysicsquestionanswering} introduced a multimodal physics dataset, expanding from 300 manually created questions to 4,500 using LLMs; \citet{liang2024scemqascientificcollegeentrance} developed SceMQA, a dataset of 1,000+ scientific reasoning problems for students transitioning to college; \citet{zhang2023m3exammultilingualmultimodalmultilevel} and \citet{das2024examsvmultidisciplinemultilingualmultimodal} created multilingual, multimodal benchmarks across various subjects and difficulty levels. 

While these benchmarks provide insight into AI capabilities, they primarily evaluate models in isolation, without comparing their performance to humans. As a result, it is unclear whether a model’s low performance stems from its limitations or if the problems themselves are inherently difficult for humans too. Understanding what makes a problem easier or harder for AI compared to humans would help warn students about potential risks and guide the design of fairer image-based assessments.

We compile 201 university-level STEM exam questions with images from Bachelor's and Master's programs across 11 subjects of varying complexity. To analyze model performance, each question is manually annotated with its image type, role, question type, and problem type. In addition, we collect student performance data, each question receiving at least five responses and an average of 546 respondents across the dataset. To evaluate AI performance, we implement five prompting strategies and test two models from GPT-family —\texttt{GPT-4o} and \texttt{o1-mini}~\citep{openai2023gpt4} as performant models freely available to students, and \texttt{Qwen 2.5 72B VL}~\citep{bai2025qwen25vltechnicalreport}, \texttt{DeepSeek r1}~\citep{deepseekai2025deepseekr1incentivizingreasoningcapability}, and \texttt{Claude 3.7 Sonnet, 2025} as performant models with visual capabilities.

%In this work, we compile 201 multimodal university-level STEM exam questions with image and text components from Bachelor’s and Master’s programs. Our data set covers 11 subjects with varying levels of complexity. Each question is manually annotated by image type, image role, question type, and problem type. In addition, we collect student performance data on these questions with a minimum of 5 responses per question and the average number of respondents of 546.  We implement five prompting strategies and evaluate two models — \texttt{GPT-4o} and \texttt{o1-mini}~\citep{openai2023gpt4}.\footnote{considered to be highly performant as of September 2024} 

Our results indicate that while LLMs perform well in text-based university assessments~\citep{borges2024}, they struggle with questions involving visual components. On average, models perform slightly worse than students. Student performance varies by subject, while model performance depends on question and image features. Based on the analysis, we provide recommendations for designing take-home assignments that maintain academic integrity by challenging models without increasing difficulty for students. These principles can also inform the development of more challenging benchmarks as models continue to improve.

\begin{figure}[t]
  \includegraphics[width=\columnwidth]{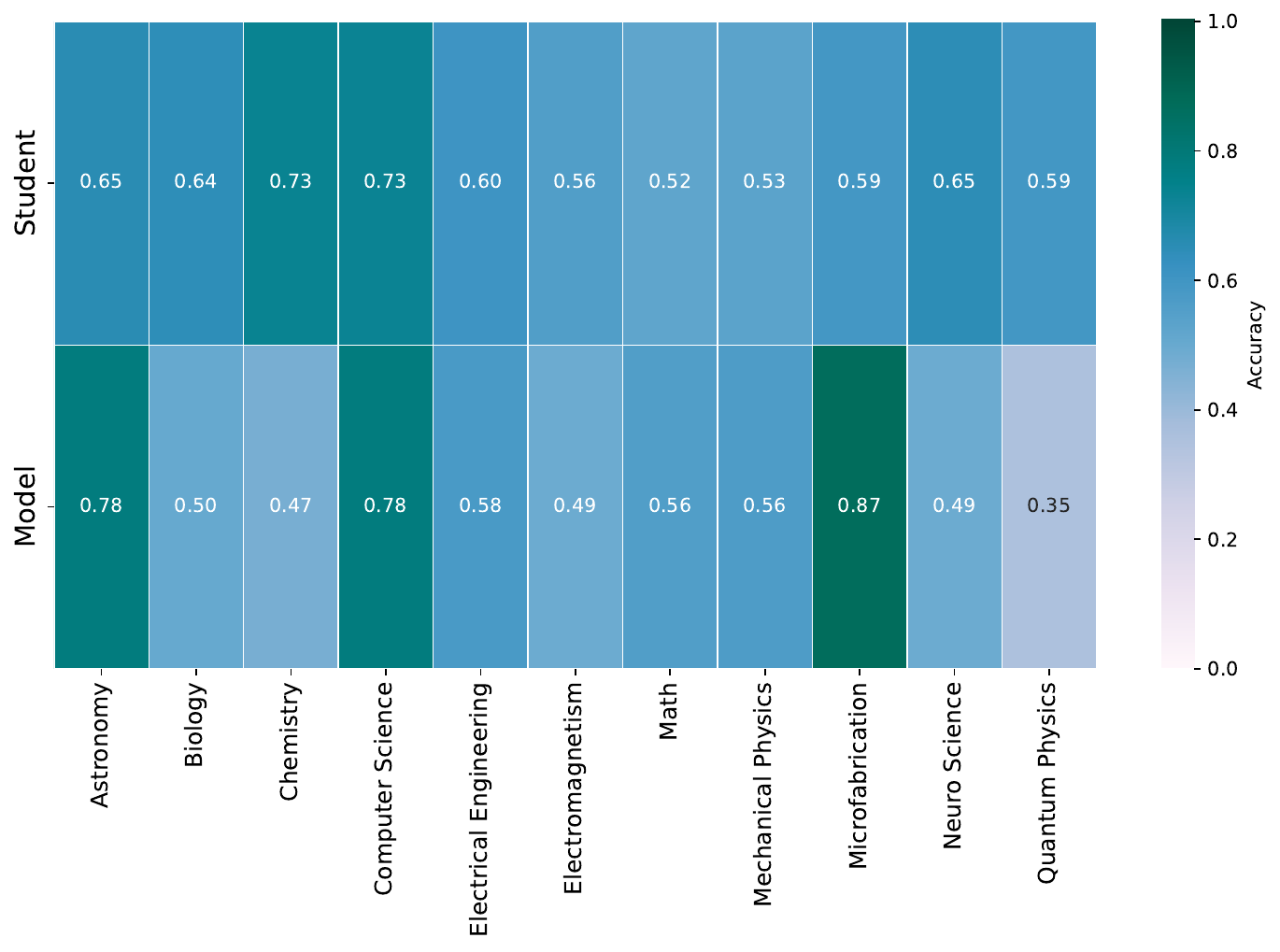}
  \caption{Average model and student accuracy per subject, with model results aggregated using the majority vote strategy.}
  \label{fig:subj}
\end{figure}

%Many benchmarks highlight in their error analysis as the reasons of failures incorrect knowledge retrieval, flawed logical reasoning, and computational errors.

\begin{figure*}
 \centering
 \begin{subfigure}[t]{.40\textwidth}
     \centering
     \includegraphics[width=\textwidth]{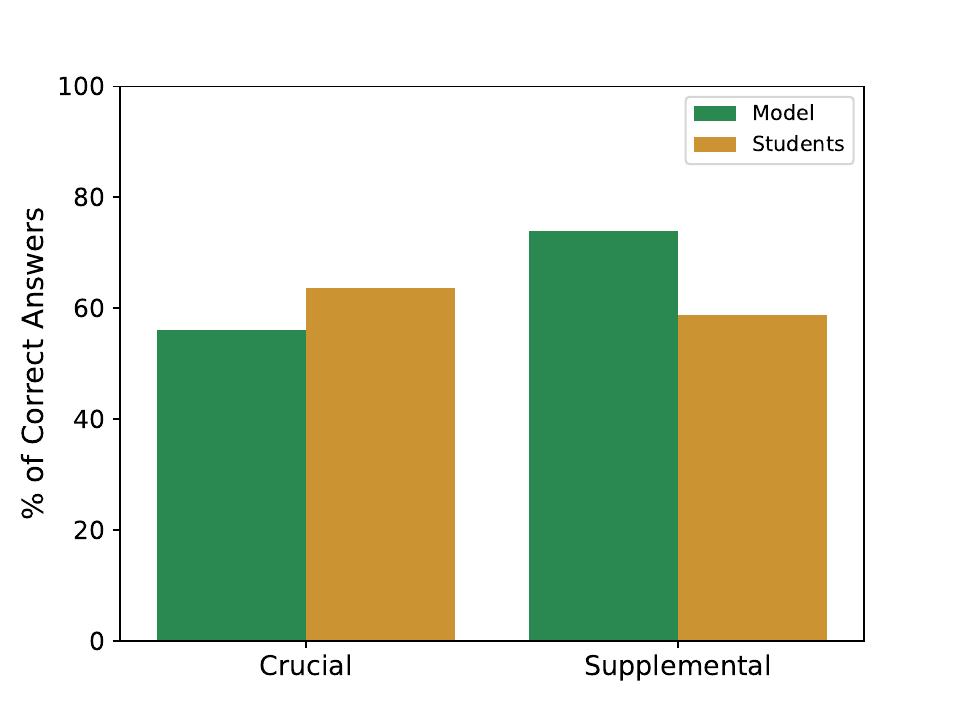}
    \caption{Image Property}\label{fig:crucial}
 \end{subfigure}
\begin{subfigure}[t]{.40\textwidth}
     \centering
     \includegraphics[width=\textwidth]{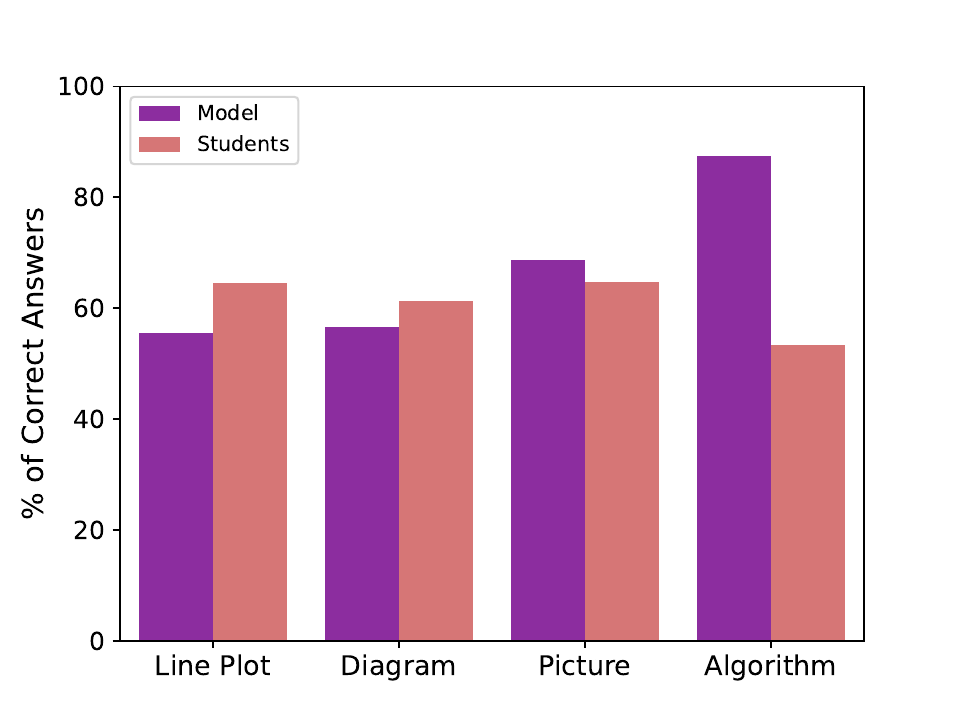}
   \caption{Image Type}\label{fig:image-type} \end{subfigure}
\caption{(a) Effect of image properties on model performance aggregated by the majority vote strategy compared to average student performance.
(b) Effect of image type on model performance aggregated by the majority vote strategy compared to average student performance.}\label{fig:allignment}
 \end{figure*}

\section{Data Set Description}

We manually collected 201 questions with images from exams and quizzes in 11 subjects from Bachelor's and Master's programs. Each question is paired with a gold answer provided by the educator who authored it. Questions were manually labeled with the following attributes\footnote{The dataset is available on \href{https://github.com/AnnaSou/multimodal-stem-ai}{GitHub}}:

\noindent \textbf{Image Type:} diagram, line plot, algorithm, and picture. \\
\noindent \textbf{Image Purpose:} An image is ``supplemental'' if all necessary information is in the text and can be inferred without it. It is ``crucial'' if required to solve the problem.\\
%An image is supplemental when it is not essential to solving the problem — all the necessary information is provided in the text and can be inferred or visualized from the description. In contrast, an image is crucial when it is required to solve the problem. \\
\noindent \textbf{Question Type:} multiple choice questions (MCQ), multiple choice questions multiple answers (MCQ-MA), and compound questions containing multiple sub-MCQ questions connected by the same question topic and having some related information in each other.\\
\noindent \textbf{Complexity of Problem Conditions:} %``Complex'' means that the question involves multiple concepts of the subject, while ``Simple'' would require only one or two closely related concepts to solve the problem. This condition does not directly reflect problem difficulty; a question may involve a single difficult concept or multiple simple ones. Distinguishing between simple and complex questions allowed us to evaluate whether models struggled with interdependent conditions. Simple questions involve fewer variables, while complex ones require integrating multiple pieces of information. 
``Complex'' questions involve multiple subject concepts, while ``Simple'' ones require only one or two closely related concepts. This distinction does not indicate difficulty—a question may have one hard concept or multiple simple ones. Categorizing questions this way helped assess whether models struggled with interdependent conditions, as simple questions have fewer variables, while complex ones require integrating more information.

Student performance data was collected from historical course records as aggregated statistics, with 5 to 5,686 respondents per question (average: 546). Student performance also served as an indicator of problem difficulty. Our dataset includes 43 problems where fewer than 40\% of the students answered correctly, 79 where 40–70\% succeeded, and 79 where more than 70\% solved the question. 

For detailed dataset statistics and student performance, see Appendices \ref{ap:dataset} and \ref{ap:student perf}.

%For details on student performance and dataset information see Appendix \ref{ap:student perf} and \ref{ap:dataset}.

% Initial experiments revealed that intricate prompts emphasizing image analysis did not significantly outperform simpler strategies. This finding suggested that ChatGPT-4o’s image processing capabilities were sufficient, but reasoning might be the bottleneck in achieving better performance. To investigate this hypothesis, we incorporated o1 mini and o1 preview into our experiments. These models were known for their strong reasoning capabilities but lacked vision support. To bridge this gap, we employed ChatGPT-4o as their "eyes" by generating detailed image descriptions for input into the o1 models. This approach necessitated a novel prompting strategy, wherein ChatGPT-4o was tasked with describing images in the context of the accompanying text to provide the most relevant information for the o1 models. This setup allowed us to isolate reasoning capabilities from vision processing. To ensure rigor, we also tested ChatGPT-4o with the same image description prompts used for o1 mini and o1 preview. This comparative approach helped us determine whether the bottleneck was primarily in the reasoning capabilities of the o1 models or a result of the differences in input processing strategies. By analyzing these setups, we gained deeper insights into the interplay between vision and reasoning in multimodal tasks.

\section{Experiments}

Our experiments assess model performance across five prompting strategies and compare it with human performance. Details on prompting strategies are provided in Appendix \ref{ap:prompting}. For multiple-choice (MCQ, MCQ-MA) and compound questions, we use exact match with the gold answer without partial credit. Model scores are aggregated using two methods: majority vote (assigning the most common score across strategies) and max (taking the highest achieved score). The max approach provides an upper bound estimate, highlighting if at least one strategy yields the correct answer. Model implementation details are in Appendix \ref{ap:model conf}.
%We evaluated GPT-4o, o1-mini, Qwen 2.5 72B VL, DeepSeek r1, and Claude 3.7 Sonnet with implementation details in Appendix \ref{ap:model conf}.

\section{Analysis}

This section presents the experimental results, comparing the model and student performance across various dimensions.

\subsection{General performance on questions with images}

Unless stated otherwise, we use majority vote aggregation. GPT-4o outperforms other models, and we focus on its results throughout. Detailed model comparisons are provided in Appendix \ref{app:model comparison}. As shown in Figure \ref{fig:method}, all prompting strategies perform similarly and roughly match the average human student’s performance on the task.

We analyze model and student accuracy on image-based questions across subjects (Figure \ref{fig:subj}). Both exhibit subject-specific strengths and weaknesses, but student accuracy varies less (0.52–0.73) than the model's (0.35–0.87). The model performs exceptionally well in Astronomy, Computer Science (CS), and Microfabrication, likely due to the structured nature of these questions and the model’s ability to apply general concepts. Prior studies have shown that LLMs excel at CS-related tasks~\citep{krüger2023performancelargelanguagemodels, song2024csbenchcomprehensivebenchmarklarge, borges2024}. In contrast, the model struggles with Quantum Physics, Chemistry, Neuroscience, and Electromagnetism, where complex, content-rich images may pose additional challenges. 
\begin{figure*}
 \centering
 \begin{subfigure}[t]{.40\textwidth}
     \centering
     \includegraphics[width=\textwidth]{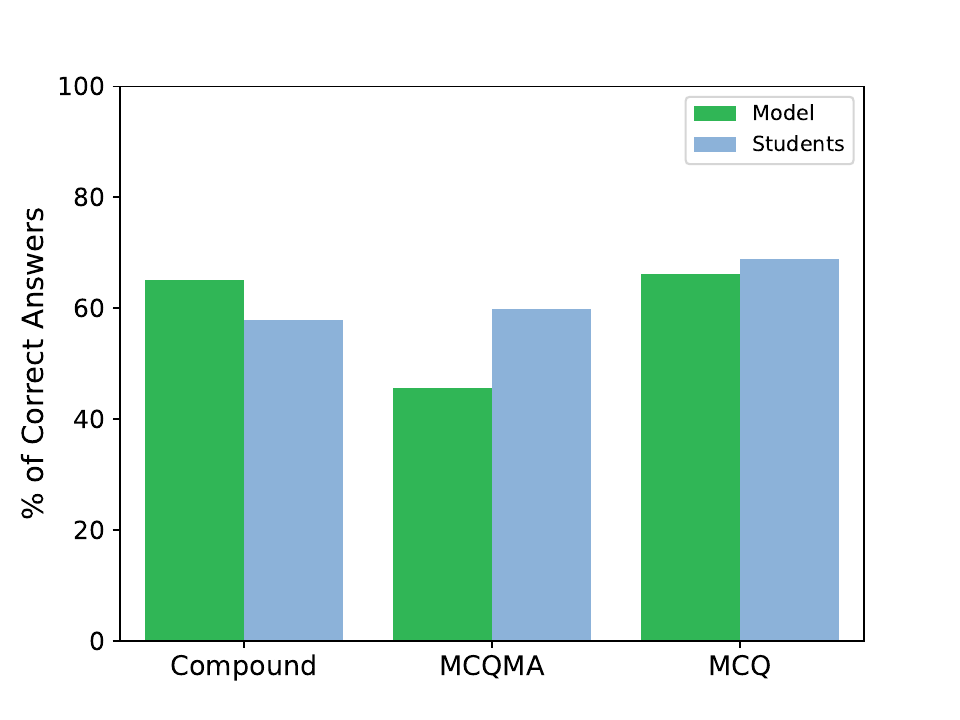}
    \caption{Question Type}\label{fig:question-type}
 \end{subfigure}
\begin{subfigure}[t]{.40\textwidth}
     \centering
     \includegraphics[width=\textwidth]{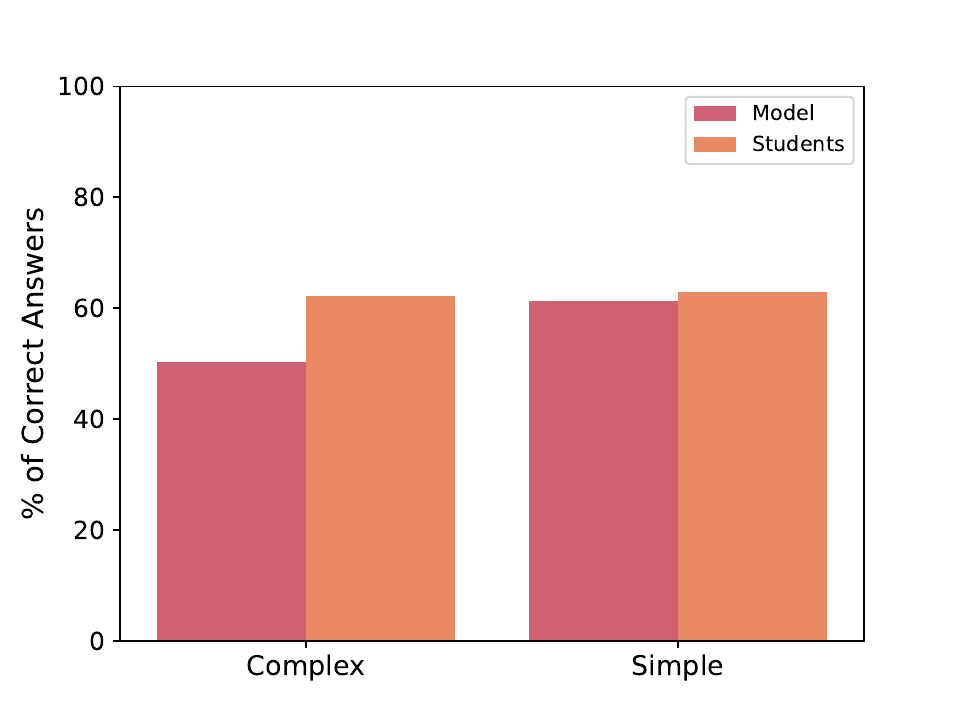}
   \caption{Problem Type}\label{fig:prob-type} \end{subfigure}
\caption{(a) Effect of question type on model performance aggregated by the majority vote strategy compared to average student performance. (b) Effect of problem type on model performance aggregated by the majority vote strategy compared to average student performance. }\label{fig:allignment2}
 \end{figure*}

\subsection{Effect of image features}

We examine the role of images in problem-solving, specifically whether they provide essential information absent from the text or if the problem can be solved without them. Figure \ref{fig:crucial} compares performance based on image necessity. As expected, student accuracy remains similar regardless of image importance, whereas models perform better on questions where images are non-essential. Our ablation study confirms this trend: removing supplemental images slightly improves model performance, though the effect is minimal (see Appendix \ref{ap:supplemental} for details).

Next, we analyze performance across image types (Figure \ref{fig:image-type}). Students perform similarly across line plots, diagrams, and pictures, and struggle the most with algorithm questions. Although the model has no difficulties in processing algorithms, it struggles the most with diagrams and line plots.

% \begin{figure}[t]
%   \includegraphics[width=\columnwidth]{figures/Crucial_supplemental.pdf}
%   \caption{Model performance aggregated by the majority vote strategy compared to average student performance per image feature.}
%   \label{fig:crucial}
% \end{figure}

% \begin{figure}[t]
%   \includegraphics[width=\columnwidth]{latex/figures/Image_type.pdf}
%   \caption{Model performance aggregated by the majority vote strategy compared to average student performance per image type. }
%   \label{fig:image-type}
% \end{figure}

\subsection{Effect of question features}

We observe that students perform similarly across all three question formats. Both students and the model get the best performance on MCQ questions. The model performs slightly better than students on compound questions, a subset of MCQs that are linked to represent steps of a larger problem. However, it struggles the most with MCQMA, often selecting some correct choices but failing to identify all (Figure \ref{fig:question-type}).

Figure \ref{fig:prob-type} illustrates how the concept count influences performance. Although students perform consistently regardless of the number of concepts in a question, the model struggles when more than two concepts are involved. 

We report statistical significance for the model's and students' results in Tables \ref{tab:accuracy_model} and \ref{tab:accuracy_student}.

\subsection{Error analysis}

%To analyze the model’s strengths and weaknesses, we identify two question sets: one where students performed poorly but the model excelled, and another where students succeeded but the model struggled. This resulted in 59 questions, which we qualitatively analyzed alongside model responses. Both sets share similarities, as they involve technical and scientific reasoning, require formula application, and follow mathematical structures.
To assess the model's strengths and weaknesses, we analyzed 59 questions split into two sets: ones where the model outperformed students and ones where it underperformed.

\noindent \textbf{Questions easy for students, hard for the model}
We examined 31 questions where the model scored 0 using the max strategy—failing to produce a correct answer across five prompts—while students achieved over 40\% accuracy. We find that humans more easily integrate common sense, domain-specific intuition, and experiential learning, whereas the model struggles to infer conditions or iterations that are not explicitly stated.

One notable category where students outperform the model involves physics-based reasoning and real-world conventions. These problems require understanding implicit relationships, precise numerical or symbolic extraction, and intuition-driven problem-solving. For instance, students effectively interpret diagrams, such as photonic crystal defects or force distributions in mechanical systems, while the model struggles with directional trends and recognizing constraints in visual data. Furthermore, the model has difficulty selecting the correct schema or plot from multiple options, a task that poses less challenge for humans.

\noindent \textbf{Questions hard for students, easy for the model}
We analyze 28 questions where students' performance is below 40\% while the model scores above 65\%.

A key category where the model outperforms students includes problems requiring structured reasoning, precise pattern recognition, and large-scale knowledge retrieval. These problems follow well-defined rules, abstract mathematical principles, and algorithmic logic. The model’s ability to detect structural patterns allows it to efficiently analyze periodicity in trigonometry, solve algorithmic network problems, and interpret simple electrical schematics with high accuracy. Unlike intuition-driven tasks, these problems follow clear logical steps. Students often struggle with multi-step reasoning due to cognitive load, whereas the model processes extended contexts effortlessly. As shown in Figure \ref{fig:char}, student accuracy declines as question length increases, while the model maintains strong performance. Additionally, models excel in problems requiring abstraction and conceptual knowledge.

\section{Conclusion}

We show that questions requiring crucial images and multiple concepts, while remaining concise, pose a greater challenge for models without increasing difficulty for students. Additionally, models struggle more than humans in applying domain-specific intuition to problem-solving. However, our analysis reveals that models retain knowledge of the correct answer in 75.5\% of questions across at least one prompting strategy but fail to retrieve it consistently. With a majority vote strategy, models achieve 58.5\% accuracy, slightly below the human average of 62.7\%. While overall performance appears similar, a closer analysis highlights the significant impact of the problem and image features on these results.

Finally, it is important to balance fair accessibility with preventing model misuse, as restrictive measures may inadvertently disadvantage students with vision impairments. For these students, problems with supplemental images are easier to understand through full-text descriptions, similar to how models rely on textual input over visual data.

\section*{Limitations}

Our study explores how humans and models solve questions involving both images and text. However, it has several limitations.

First, our dataset is relatively small (201 examples). While we ensured high-quality data through manual collection and annotation and confirmed statistical significance, a larger dataset would improve reliability. We opted against automated data augmentation to maintain quality control. To facilitate further research, we publicly release our dataset with annotations.

Second, our grading method does not assign partial credit for multiple-choice multiple-answer (MCQMA) questions, leading to a stricter evaluation of model performance. Additionally, unlike humans, models do not employ elimination reasoning, as we do not adjust prompts for MCQMA responses, potentially disadvantaging them.

Third, when comparing course performance, we do not account for instructor influence, which may affect problem difficulty. This factor can introduce bias also for humans, as different instructors may present varying challenges for students within the same subject.

\section*{Acknowledgments}
We are grateful to Dr. Jessica Dehler Zufferey for her valuable recommendations on question and image feature design, as well as her feedback on the interpretation of our results. We also thank Christian Vonarburg and Yves Renier for their assistance with data preparation. We appreciate the feedback provided by the EPFL NLP lab members throughout the development of this work. We also gratefully acknowledge the support of the Swiss National Science Foundation (No. 215390), Innosuisse (PFFS-21-29), the EPFL Center for Imaging, Sony Group Corporation, and a Meta LLM Evaluation Research Grant.

\bibliography{custom}

\appendix

\section{Data Set Details}
Here we present the data set statistics, and examples of questions for every data feature. 

\subsection{Data Format} \label{ap:dataset}

\begin{figure}[t]
  \includegraphics[width=\columnwidth]{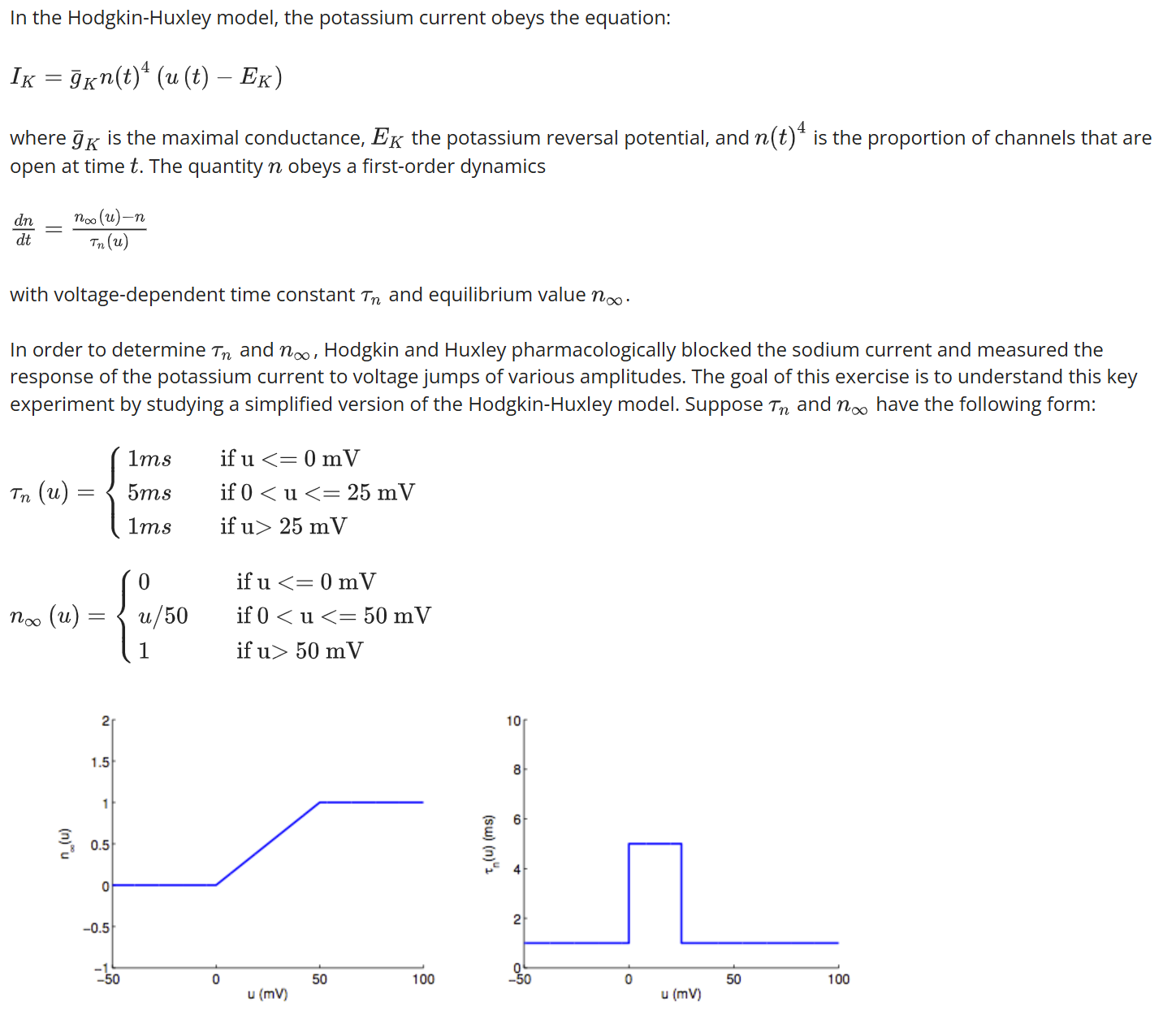}
  \caption{Example of a complex question with supplemental image.}
  \label{ap:fig:example_complex_supplemental}
\end{figure}

\noindent \textbf{Image Type:} diagram, line plot, algorithm, and picture. \\
\noindent \textbf{Image Purpose:} supplemental, when the image is non-essential and all information about the problem is stated in the problem text or crucial, when the image is required to solve the problem. One can determine that the question in Figure \ref{ap:fig:example_complex_supplemental} has a supplemental image since it could be inferred from the text. \\

\noindent \textbf{Question Type:} multiple choice questions (MCQ), multiple choice questions multiple answers (MCQ-MA), and compound questions containing multiple sub-MCQ questions connected by the same question topic and having some related information in each other.\\
\noindent \textbf{Complexity of Problem Conditions:} ``Complex'' means that the question involves multiple concepts of the subject, while ``Simple'' would require only one or two closely related concepts to solve the problem. This condition does not directly reflect problem difficulty; a question may involve a single difficult concept or multiple simple ones. Distinguishing between simple and complex questions allowed us to evaluate whether models struggled with interdependent conditions. Simple questions involve fewer variables, while complex ones require integrating multiple pieces of information. 

One can determine that the question in Figure \ref{ap:fig:example_complex_supplemental} is ``Complex'' and the image is ``Supplemental''. The question is complex because it involves the Hodgkin-Huxley model, differential equations governing potassium channel dynamics, and voltage-dependent parameters, requiring knowledge of electrophysiology and mathematical modeling. The image is supplemental because it provides graphical representations of \( n_{\infty}(u) \) and \( \tau_n(u) \), but all necessary equations and definitions are clearly described in the text, making the image helpful but not essential.

\subsubsection{Description of Labels}
\label{sec:label_description}
\begin{enumerate}
    \item \textbf{Course\_name}:
    \begin{itemize}
        \item \textit{Description}: The name or identifier of the course associated with the question.
        \item \textit{Example}: \texttt{"Calculus I"}, \texttt{"Physics 101"}
    \end{itemize}
    
    \item \textbf{Exercise\_name}:
    \begin{itemize}
        \item \textit{Description}: The unique exercise id.
    \end{itemize}

    \item \textbf{Question}:
    \begin{itemize}
        \item \textit{Description}: The text of the question, may include LaTeX formatting and placeholders for images.
    \end{itemize}

    \item \textbf{Gold\_answer}:
    \begin{itemize}
        \item \textit{Description}: The correct answer to the question. 
    \end{itemize}

    \item \textbf{Question\_type}:
    \begin{itemize}
        \item \textit{Description}: The format or type of the question.
        \item \textit{Possible Labels}:
        \begin{itemize}
            \item \texttt{"MCQ"} (Multiple Choice Question)
          %  \item \texttt{"Open-ended"}
            \item \texttt{"MCQMA"} (MCQ Multiple Answers)
            \item \texttt{"Compound"} (a non-open-ended question with multiple objectives)
        \end{itemize}
    \end{itemize}

    \item \textbf{Image\_type}:
    \begin{itemize}
        \item \textit{Description}: The type of images included in the question.
        \item \textit{Possible Labels (Others may be added as we manually label)}:
        \begin{itemize}
            \item \texttt{"line plot"}
            \item \texttt{"bar plot"}
            \item \texttt{"scatter plot"}
            \item \texttt{"histogram"}
            \item \texttt{"pie chart"}
            \item \texttt{"table"}
            \item \texttt{"image"}
            \item \texttt{"diagram"}
        \end{itemize}
    \end{itemize}

    \item \textbf{Image\_purpose}:
    \begin{itemize}
        \item \textit{Description}: The role of the image in the context of the question.
        \item \textit{Possible Labels}:
        \begin{itemize}
            \item \texttt{"Crucial"} (Essential for solving the question)
            \item \texttt{"Supplemental"} (Doesn’t provide additional context)
        \end{itemize}
    \end{itemize}

    \item \textbf{Problem\_conditions}:
    \begin{itemize}
        \item \textit{Description}: The complexity of the conditions within the problem.
        \item \texttt{"Complex"} doesn’t necessarily mean that the problem is difficult. It simply means that many conditions are in play.
        \item \textit{Possible Labels}:
        \begin{itemize}
            \item \texttt{"Simple"} (Conditions are straightforward and not interacting)
            \item \texttt{"Complex"} (Multiple conditions interact to find the answer)
        \end{itemize}
    \end{itemize}

    \item \textbf{Question\_images}:
    \begin{itemize}
        \item \textit{Description}: A list of filenames or identifiers for images included in the question.
    \end{itemize}

    \item \textbf{Question\_length\_characters}:
    \begin{itemize}
        \item \textit{Description}: The length of the question text is measured in characters.
    \end{itemize}

    \item \textbf{Num\_objectives}:
    \begin{itemize}
        \item \textit{Description}: The number of sub-questions within the question.
        \item \textit{Example}: \texttt{1}, \texttt{2}
    \end{itemize}

    \item \textbf{Language}:
    \begin{itemize}
        \item \textit{Description}: The language in which the question is written.
        \item Always in \texttt{"English"} because we translated the French ones.
    \end{itemize}

    \item \textbf{Original\_language}:
    \begin{itemize}
        \item \textit{Description}: The original language of the question before translation.
        \item \textit{Example}: \texttt{"French"}
    \end{itemize}

    \item \textbf{Was\_translated}:
    \begin{itemize}
        \item \textit{Description}: Indicates whether the question was translated from another language.
        \item \textit{Possible Values}: \texttt{true} or \texttt{false}
    \end{itemize}

    \item \textbf{Image\_file\_type}:
    \begin{itemize}
        \item \textit{Description}: The file format of the images used.
        \item \textit{Example}: \texttt{"PNG"}, \texttt{"JPEG"}
    \end{itemize}

    \item \textbf{Answer\_format}:
    \begin{itemize}
        \item \textit{Description}: The expected format of the answer.
        \item \textit{Possible Labels}:
        \begin{itemize}
            \item \texttt{"Only MCQ Letter"} (previously called MCQ)
            \item \texttt{"Only Numeric Answer"}
            \item \texttt{"Derivation"}
            \item \texttt{"Text"}
            \item \texttt{"Code"}
            \item \texttt{"Calculation"}
        \end{itemize}
    \end{itemize}

    \item \textbf{Solution\_type}:
    \begin{itemize}
        \item \textit{Description}: Indicates whether the question has a unique correct answer or multiple correct answers.
        \item \textit{Possible Labels}:
        \begin{itemize}
            \item \texttt{"Unique answer"}
            \item \texttt{"Multiple answers"}
        \end{itemize}
    \end{itemize}

    \item \textbf{Type\_of\_text}:
    \begin{itemize}
        \item \textit{Description}: The formatting or typesetting used in the question text.
        \item \textit{Example}: \texttt{"LaTeX"}, \texttt{"Plain text"}, \texttt{"XML"}
    \end{itemize}

    \item \textbf{Objective\_dependency}:
    \begin{itemize}
        \item \textit{Description}: Indicates whether the objectives in the question are independent or dependent on previous ones.
        \item \textit{Possible Labels}:
        \begin{itemize}
            \item \texttt{"All Independent"} (Objectives can be solved separately)
            \item \texttt{"Dependent"} (Some objectives rely on answers from previous parts)
        \end{itemize}
    \end{itemize}

\end{enumerate}

\subsection{Data Set Statistics}\label{ap:data stat}
Data set statistic is presented in Table \ref{ap:data stat}.

\begin{table}[ht]
\centering
\begin{tabular}{|l|c|c|}
\hline
\textbf{Feature} & \textbf{Options} & \textbf{Count} \\
\hline
\multirow{4}{*}{\makecell[l]{Question \\ Type}} 
& MCQ     & 80 \\
& Compound       & 59 \\
& MCQMA          & 46 \\
& Numeric and Formula & 16 \\
\hline

\multirow{5}{*}{\makecell[l]{Image \\ Type}}  
& Diagram        & 109 \\
& Line Plot      & 45 \\
& Picture        & 33 \\
& Algorithm      & 8 \\
& Other          & 6 \\
\hline

\multirow{2}{*}{\makecell[l]{Image \\ Purpose}}  
& Crucial        & 162 \\
& Supplemental   & 39 \\
\hline

\multirow{2}{*}{\makecell[l]{Problem \\ Conditions}}  
& Simple         & 169 \\
& Complex        & 32 \\
\hline

\multirow{11}{*}{\makecell[l]{Course \\ Category}}  
& Astronomy             & 32 \\
& Electrical Engineering & 28 \\
& Computer Science       & 20 \\
& Math                   & 19 \\
& Electromagnetism       & 17 \\
& Quantum Physics        & 26 \\
& Mechanical Physics     & 16 \\
& Neuroscience           & 15 \\
& Microfabrication       & 10 \\
& Chemistry              & 10 \\
& Biology                & 8 \\
\hline
\end{tabular}
\caption{Data Statistics}\label{ap:data stat}
\end{table}

\subsection{Student Performance}\label{ap:student perf}

\begin{figure}[t]
  \includegraphics[width=\columnwidth]{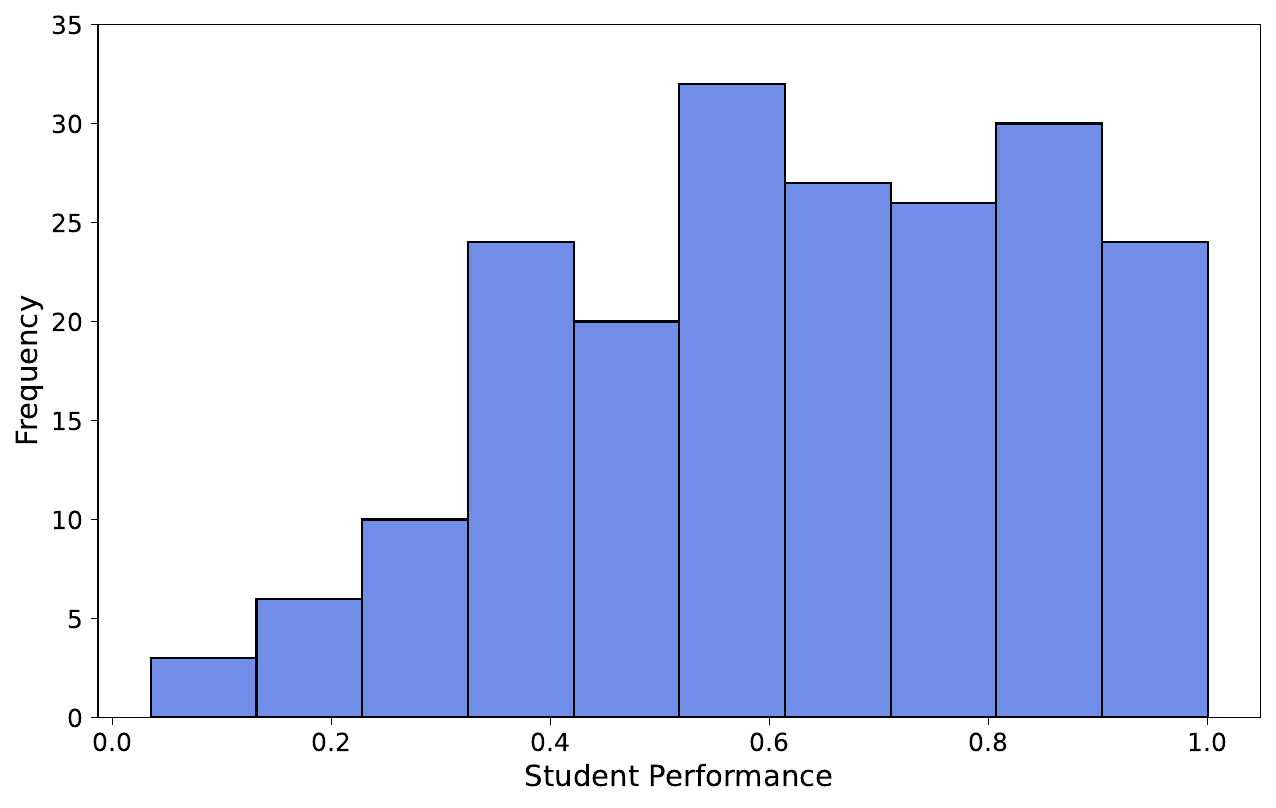}
  \caption{Student accuracy distribution.}
  \label{ap:fig:student}
\end{figure}

Dataset difficulty illustrated by student performance is presented in Figure \ref{ap:fig:student}.

The distribution of students attempting a question is presented in Table \ref{tab:respondents}.

\begin{table}[ht]
\centering
\begin{tabular}{|c|c|}
\hline
\textbf{Respondents} & \textbf{Questions} \\
\hline
5-20     &    27\\
21-50     &   34 \\
51-100    &   29 \\
101-500   &   51 \\
501-1000   &  33 \\
1001-2000  &  21 \\
2001-7000  &   6 \\

\hline
\end{tabular}
\caption{Distribution of questions by the number of respondents.}
\label{tab:respondents}
\end{table}

\section{Prompting strategies} \label{ap:prompting}

\subsection{Helper Functions}

\subsubsection{\texttt{question\_type\_prompt}}

The \texttt{question\_type\_prompt} function creates a tailored instruction based on the type of question being posed. It supports several question types, each associated with a specific directive:
\begin{itemize}
    \item \textbf{MCQ:} Instructs the model to select the correct option by returning only its letter.
    \item \textbf{MCQMA:} Similar to MCQ but expects multiple correct options, concatenated as a single string (e.g., \texttt{AB} rather than \texttt{A, B}).
    \item \textbf{Numeric Question:} Requests that the model output only the numerical answer.
    \item \textbf{Formula Question:} Expects the answer to be provided as a formula.
    \item \textbf{Open Ended:} Directs the model to comprehensively address all parts of the question.
\end{itemize}

For questions labeled as \textbf{Compound}, the function combines the individual instructions corresponding to each subquestion type. It first determines the number of subquestions and then appends the respective prompt text for each, ultimately guiding the model to return its answers as a JSON-formatted list.

\vspace{1em}

\subsubsection{\texttt{generate\_format\_instruction}}

The \texttt{generate\_format\_instruction} function provides context-specific formatting advice based on the text's format:
\begin{itemize}
    \item \textbf{XML:} The instruction reminds the model to interpret XML symbols correctly, ensuring that any formula or question components formatted in XML are properly understood.
    \item \textbf{LaTeX:} Advises careful interpretation of LaTeX expressions, especially for mathematical content.
    \item \textbf{Other:} When the text does not fall into the above categories, no extra formatting instruction is provided.
\end{itemize}

\subsection{Prompting strategies}

To generate question-answer pairs, we first conducted an experiment evaluating 12 different prompting strategies. Based on performance results, we selected five strategies for further analysis. Two of these serve as baselines: \textit{direct zero-shot}, the model receives only the question and image without additional instructions or contextual information. \textit{zero-shot chain-of-thought (CoT)}~\citep{wei2023chainofthought}, the model is asked to produce intermediate reasoning steps before arriving to the final answer. Beyond the baselines, we investigated how the order of multimodal input affects performance. Specifically, we compared cases where the model processes the image at the beginning versus at the end of the text input. Our results indicate that presenting the \textit{image first}, followed by the problem text, leads to better performance. Finally, for models with strong reasoning capabilities but lacking the multimodal component, we implemented a \textit{two-stage} prompting strategy. We first use GPT-4o to generate a textual description of the image. This description is then passed, along with the problem text, to o1-mini and o1-preview models. The quality of the generated descriptions were manually verified.

\subsubsection{Direct zero-shot}
A straightforward prompt that presents the question to the model.
\begin{tcolorbox}[colback=promptblue, colframe=promptblue!80!black, title=Direct zero-shot]
\small
You are an expert in STEM courses.\\

Images:  
\textless image\_names \textgreater \\

[Refer to \texttt{generate\_format\_instruction}]

Question: \textless question text \textgreater\\

[Refer to \texttt{question\_type\_prompt}]  \\
Your Answer:
\end{tcolorbox}

\subsubsection{Chain-of-Thought Prompt}
This prompt encourages a step-by-step analytical approach, asking the model to think through the problem before answering.
\begin{tcolorbox}[colback=promptgreen, colframe=promptgreen!80!black, title= Chain-of-Thought Prompt]
\small
You are an expert in STEM courses tasked with answering questions with step-by-step analysis.

Examine both the image(s) and question text before answering.\\  

Images:  
\textless image\_names \textgreater \\

[Refer to \texttt{generate\_format\_instruction}]  \\
Question: \textless question text \textgreater \\

[Refer to \texttt{question\_type\_prompt}]\\  
Your Answer: Let's think step by step.
\end{tcolorbox}

\subsubsection{Image First Prompt}
This prompt prioritizes image analysis by instructing the model to examine the image details before considering the text, and then synthesize a detailed answer.
\begin{tcolorbox}[colback=promptyellow, colframe=promptyellow!80!black, title=Image-First Prompt]
\small
You are an expert in STEM courses tasked with answering questions. But, first, you must analyze the image(s), which you will follow with the textual analysis.\\

You will follow the next steps before providing an answer.\\
Step 1: Analyze the Image(s) First\\
\quad - Describe elements, patterns, and relationships in the image(s).\\
Step 2: Use Observations to Analyze the Text\\
\quad - Use the image understanding to find relevant textual information in the question.\\
Step 3: Provide a Detailed Answer\\
\quad - Synthesize observations into a complete answer.\\

Images:  
\textless image\_names \textgreater \\

[Refer to \texttt{generate\_format\_instruction}]  \\
Question: \textless question text \textgreater \\

[Refer to \texttt{question\_type\_prompt}]  \\
Make sure to tackle every step mentioned above, before you answer.\\
Your Answer:
\end{tcolorbox}

\subsubsection{Two Stage Prompt}
\subsubsection*{Image Description Prompt}
This prompt requests a detailed description of the provided image, linking its elements to the question context for use by another model. It does not answer the question but aims to provide details that will enable another to answer it.
\begin{tcolorbox}[colback=promptred, colframe=promptred!80!black, title=Image Description Prompt]
\small
I am going to provide you with a question with an image. I need you to describe this image in as many details as possible and link those details to the question and its context.\\

I will then share this description of the image with an LLM which doesn't have vision capabilities, but better reasoning skills than you. In other words, you will be the eyes for that second model. As such, it is primordial that you don't leave out any details!\\

Note that some details that you think might be useless, may not be, as such make sure that you focus on every aspect.\\

Here is the Image:  
\textless image\_names \textgreater \\

Here is the question:  
\textless question text \textgreater \\

You may now provide your detailed description. Make sure to follow the instructions that were given to you.
\end{tcolorbox}

\subsubsection*{Answer With Image Description Prompt}
This prompt asks the model to answer a question based solely on an image description, with a reference to the detailed image description provided earlier.
\begin{tcolorbox}[colback=promptpurple, colframe=promptpurple!80!black, title=Answer With Image Description Prompt]
\small
You are an expert in STEM courses and will answer a question that includes an image description.. \\

Here is the description of the image:  \\
\textless detailed image description \textgreater \\

[Refer to \texttt{generate\_format\_instruction}]  \\
Here is the question that you need to answer:  
\textless question text \textgreater \\

[Refer to \texttt{question\_type\_prompt}]  \\
Please, explain the solution and answer in the following format:
\begin{verbatim}
{
    "reasoning": "Your explanation.",
    "answer": "Your answer and nothing more."
}
\end{verbatim}
Your Reasoning and Answer:
\end{tcolorbox}

\subsection{Selecting prompting strategies}

\begin{figure}[t]
  \includegraphics[width=\columnwidth]{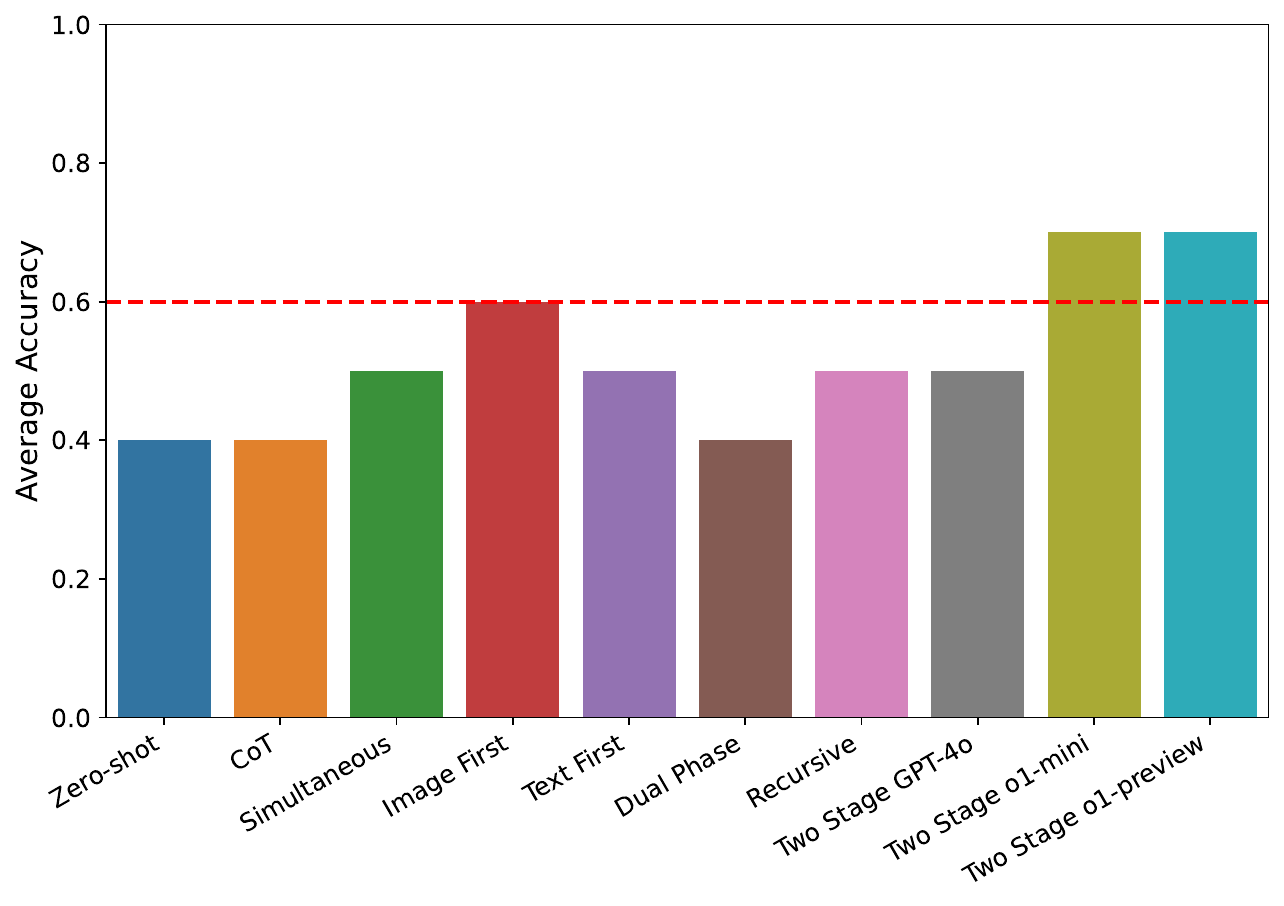}
  \caption{Model performance on the initial set of prompting strategies.}
  \label{ap:10prompts}
\end{figure}

Initially, we tested 12 prompting strategies on 10 questions to select the most effective ones for the subsequent experiments. Figure \ref{ap:10prompts} shows a comparison across all strategies. We selected the \textit{two-stage} strategy as the most effective, followed by two baseline strategies, and finally the best strategy for presenting a model with both text and image.

In the basic prompting category, the question was presented along with the image, allowing the models to interpret the visual data without additional instructions. In the second category, prompts directed the models to explicitly consider both the image and text, either together or sequentially, with varying emphasis on fine-grained versus coarse-grained details. Finally, in the third category, models lacking vision capabilities were provided with detailed descriptions of the image instead.

\subsubsection{Simultaneous Prompt }
This prompt asks the LLM to examine both image and text simultaneously, integrating insights from both modalities before answering. It emphasizes a holistic analysis that considers all available information concurrently.

\begin{tcolorbox}[colback=promptgray, colframe=promptgray!80!black, title=Simultaneous Prompt]
\small
You are an expert in STEM courses tasked with answering questions. Examine both the image(s) and question text before answering. \\

You will follow the next steps before providing an answer. \\
Step 1: Analyze the Image(s) and Text Together \\
\quad - Describe key elements, patterns, and relationships, integrating both sources. \\
Step 2: Provide a Detailed Answer\\
\quad - Synthesize observations into a complete answer. \\

Images:
\textless image\_names \textgreater \\

[Refer to \texttt{generate\_format\_instruction}] \\
Question: \textless question text \textgreater \\

[Refer to \texttt{question\_type\_prompt}] \\
Make sure to tackle every step mentioned above, before you answer.  \\
Your Answer:
\end{tcolorbox}

\subsubsection{Text First Prompt}
This prompt directs the LLM to analyze the question text initially and then examine the associated image, using the textual understanding to guide the image analysis. It ultimately expects the model to merge both insights into a coherent, well-informed answer.

\begin{tcolorbox}[colback=promptgray, colframe=promptgray!80!black, title=Text First Prompt]
\small
You are an expert in STEM courses tasked with answering questions. But, first, you must analyze the text, which you will follow with the image analysis. \\

You will follow the next steps before providing an answer.\\
Step 1: Analyze the Question Text First\\
\quad - Understand the question context.\\
Step 2:  Use Observations to Analyze the Image \\
\quad - Use textual understanding to find relevant visual information (elements, patterns, relationships, etc.)\\
Step 3: Provide a Detailed Answer \\
\quad - Synthesize observations into a complete answer. \\

Images:
\textless image\_names \textgreater \\

[Refer to \texttt{generate\_format\_instruction}] \\
Question: \textless question text \textgreater \\

[Refer to \texttt{question\_type\_prompt}] \\
Make sure to tackle every step mentioned above, before you answer. \\
Your Answer:
\end{tcolorbox}

\subsubsection{Dual Phase Prompt}
This prompt divides the analysis into two distinct phases; first analyzing the image(s) and then the text, before synthesizing the information into a final answer. It ensures that each component is evaluated independently before being combined for a comprehensive response.

\begin{tcolorbox}[colback=promptgray, colframe=promptgray!80!black, title=Dual Phase Prompt]
\small
You are an expert in STEM courses tasked with answering questions with a dual-phase approach. \\

You will follow the next steps before providing an answer. \\
Step 1: Analyze the Image(s) First\\
\quad - Describe elements, patterns, and relationships in the image(s).\\
Step 2: Interpret the Question Text Separately\\
\quad - Identify question context independently of your image findings.\\
Step 3: Synthesize Textual and Visual Information\\
\quad - Combine insights from both phases.\\
Step 4: Provide a Detailed Answer\\
\quad - Synthesize observations into a complete answer. \\

Images:
\textless image\_names \textgreater \\

[Refer to \texttt{generate\_format\_instruction}] \\
Question: \textless question text \textgreater \\

[Refer to \texttt{question\_type\_prompt}]\\
Make sure to tackle every step mentioned above, before you answer. \\
Your Answer:
\end{tcolorbox}

\subsubsection{Recursive Prompt}
This prompt directs the LLM to iteratively alternate between image and text analysis, refining its understanding with each pass until a complete picture is achieved. It is designed to produce a well-considered final answer by progressively integrating and re-evaluating both modalities.

\begin{tcolorbox}[colback=promptgray, colframe=promptgray!80!black, title=Recursive Prompt]
\small
You are an expert in STEM courses tasked with answering questions with recursive analysis. \\

You will follow the next steps before providing an answer.\\
Step 1: Analyze the Image(s) First\\
\quad - Describe elements, patterns, and relationships in the image(s).\\
Step 2: Use Observations to Analyze the Text\\
\quad - Use the image understanding to find relevant textual information in the question.\\
Step 3: Refine Analysis\\
\quad - Alternate between image and text analysis, refining observations with each pass until a comprehensive understanding of the text and image is reached.\\
Step 4: Provide a Detailed Answer\\
\quad - Synthesize observations into a complete answer.\\

Images:
\textless image\_names \textgreater \\

[Refer to \texttt{generate\_format\_instruction}] \\
Question: \textless question text \textgreater \\

[Refer to \texttt{question\_type\_prompt}] \\
Make sure to tackle every step mentioned above, before you answer. \\
Your Answer:
\end{tcolorbox}

\section{Model Configuration}\label{ap:model conf}
To evaluate performance, three OpenAI models were employed: GPT-4o with temperature 0.1, o1-mini-2024-09-12, and o1-preview-2024-09-12. GPT-4o was chosen as the baseline due to its strong vision capabilities. The o1-mini and o1-preview models, in contrast, lack native vision capabilities but exhibit strong reasoning abilities in text-based tasks. While GPT-4o allowed temperature adjustments, the o1 models did not support this feature. The primary focus was on GPT-family models as the ones that students can easily access models to run questions themselves while preparing a take-home assignment. We are trying to provide also some recommendations for educators on how to make such assignments less vulnerable to generative AI use. To explore the effect on different model families, we also test  Qwen 2.5 72B VL, r1 Deepseek, and Claude 3.7 Sonnet, 2025. 

We used \texttt{pandas},\footnote{\url{https://pandas.pydata.org/docs/index.html}} \texttt{json},\footnote{\url{https://docs.python.org/3/library/json.html}}, \texttt{numpy},\footnote{\url{https://numpy.org/doc/stable/index.html}} and \texttt{scikit-learn}\footnote{\url{https://scikit-learn.org/stable/}} to process our results, compute accuracy scores, and compute statistical significance.

\section{Additional Experimental Results}

\subsection{Model performance comparison}\label{app:model comparison}
We observe that the GPT model is the most performant one and that, in general, the models follow our findings. The results in various characteristics are presented in Table \ref{tab:accuracy_models}. 

Looking at the performance per course in Table \ref{tab:accuracy_course}, we see that our findings hold. Also, sometimes, there are cases when Claude 3.7 or R1 outperform GPT model: in a subject like biology and mechanical physics.

\begin{table*}[t!]
\centering
\small % Smaller than \small, but larger than \scriptsize
\begin{tabular}{|c|c|c|c|c|}
\hline
\textbf{Course Category} & \textbf{GPT-4o} & \textbf{Claude 3.7} & \textbf{R1} & \textbf{Qwen 2.5-72B} \\
\hline
Astronomy              & \textbf{0.78} & 0.55 & 0.58 & 0.63 \\
Biology                & 0.50 & \textbf{0.75} & 0.63 & 0.50 \\
Chemistry              & \textbf{0.47} & \textbf{0.47} & 0.37 & 0.30 \\
Computer Science       & \textbf{0.78} & 0.68 & 0.74 & 0.72 \\
Electrical Engineering & \textbf{0.58} & 0.48 & 0.37 & 0.43 \\
Electromagnetism       & \textbf{0.49} & 0.45 & 0.45 & 0.45 \\
Math                   & 0.56 & \textbf{0.57} & 0.47 & 0.56 \\
Mechanical Physics     & 0.56 & 0.56 & \textbf{0.63} & 0.56 \\
Microfabrication       & \textbf{0.87} & 0.60 & 0.64 & 0.52 \\
Neuro Science          & \textbf{0.49} & 0.31 & 0.29 & 0.36 \\
Quantum Physics        & \textbf{0.35} & 0.29 & 0.18 & 0.24 \\
\hline
\end{tabular}
\caption{Average model performance across different course categories.}
\label{tab:accuracy_course}
\end{table*}

\begin{table*}[t!]
\centering
\normalsize % Use \scriptsize if you need it even smaller
\begin{tabular}{|c|c|c|c|c|c|}
\hline
\textbf{Category} & \textbf{Label} & \textbf{GPT-4o} & \textbf{Claude 3.7} & \textbf{R1} & \textbf{Qwen 2.5-72 B} \\
\hline
\multirow{5}{*}{\textbf{Question feature}} 
  & Simple       & \textbf{0.613} & 0.517 & 0.469 & 0.481 \\
  & Complex      & \textbf{0.503} & 0.454 & 0.459 & 0.502 \\
  & MCQ          & \textbf{0.663} & 0.563 & 0.538 & 0.575 \\
  & MCQMA        & \textbf{0.457} & 0.370 & 0.304 & 0.283 \\
  & Compound     & \textbf{0.650} & 0.566 & 0.533 & 0.538 \\
\hline
\multirow{2}{*}{\textbf{Image feature}} 
  & Crucial      & \textbf{0.561} & 0.473 & 0.426 & 0.448 \\
  & Supplemental & \textbf{0.740} & 0.652 & 0.641 & 0.635 \\
\hline
\end{tabular}
\caption{Average model performance across different question and image features.}
\label{tab:accuracy_models}
\end{table*}

\subsection{Removing supplemental image}\label{ap:supplemental}

\begin{figure}[t]
  \includegraphics[width=\columnwidth]{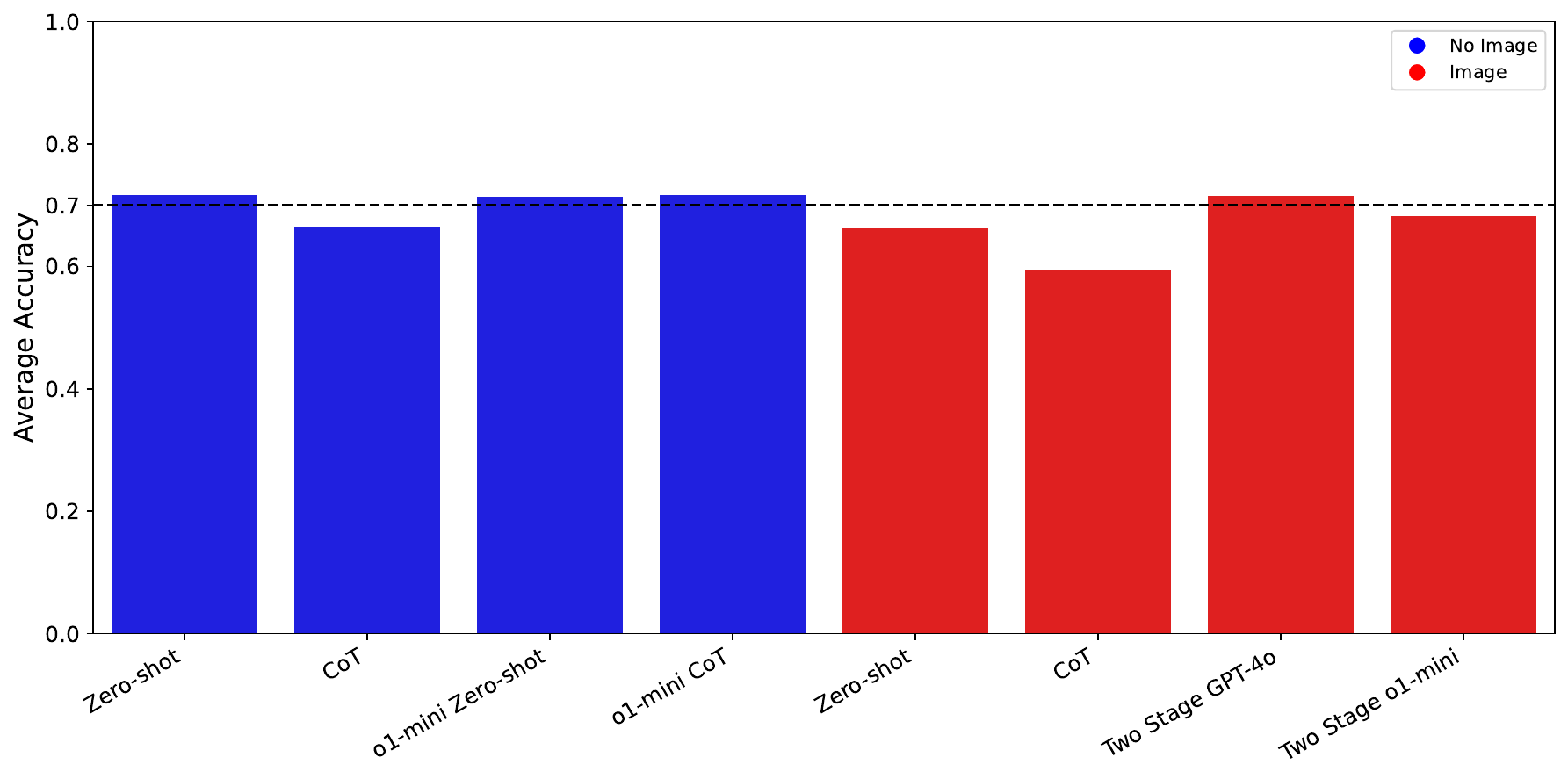}
  \caption{Model performance with and without supplemental image included.}
  \label{ap:supp}
\end{figure}

We tested the same prompts with and without supplemental images. For the \textit{two stage} prompts we removed mentions of the image and didn't pass the descriptions. Figure \ref{ap:supp} shows that the presence or absence of the image doesn't affect model performance.

\subsection{Model performance vs student performance}

\begin{figure}[t]
  \includegraphics[width=\columnwidth]{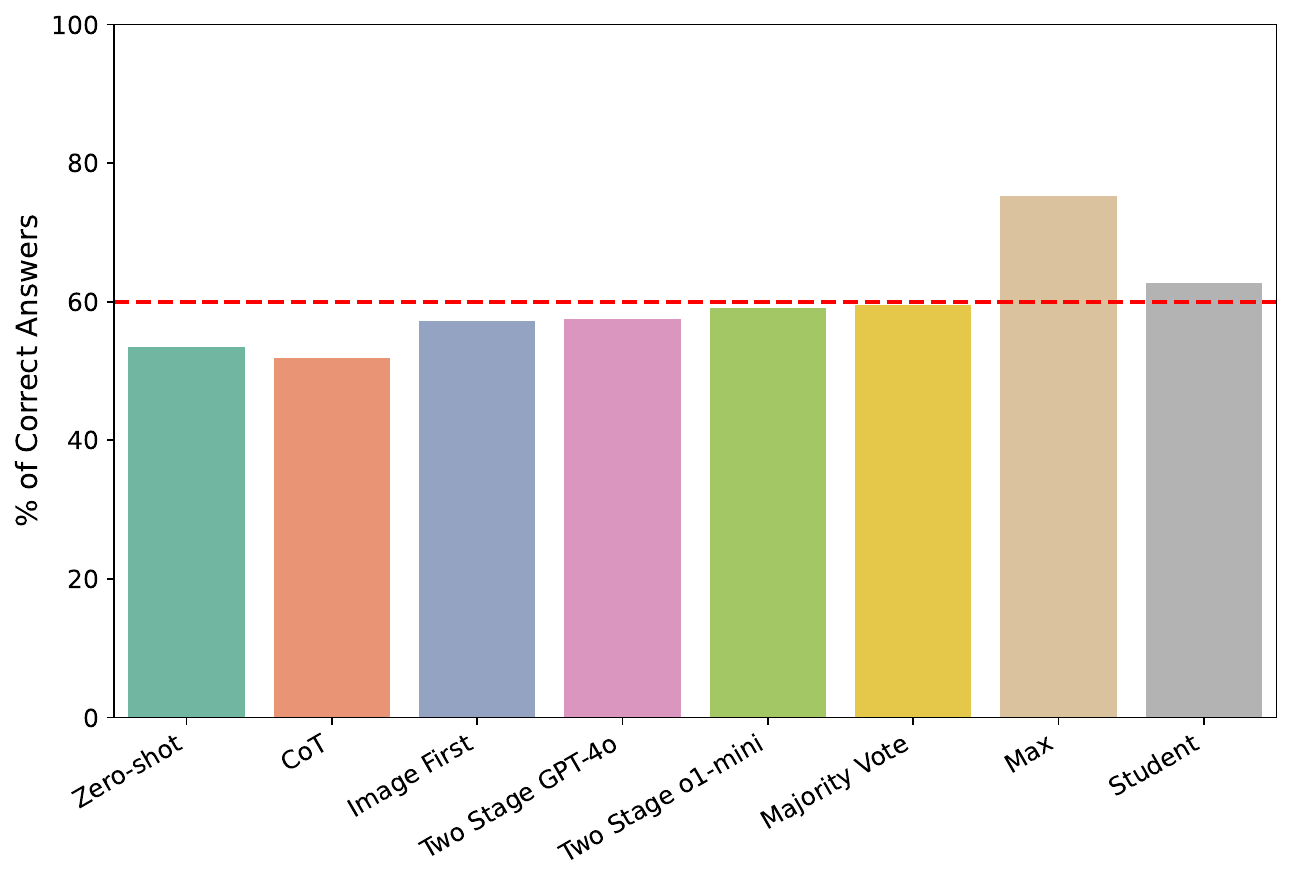}
  \caption{Average GPT-family models performance across five prompting strategies, aggregated results with the majority vote and maximum strategy and student performance. }
  \label{fig:method}
\end{figure}

In Figure \ref{fig:method}, we compare model performance across five prompting strategies and two aggregation strategies with average student performance.

\subsection{Student vs model accuracy depending on the question length}\label{ap:question length}

\begin{figure}[t]
  \includegraphics[width=\columnwidth]{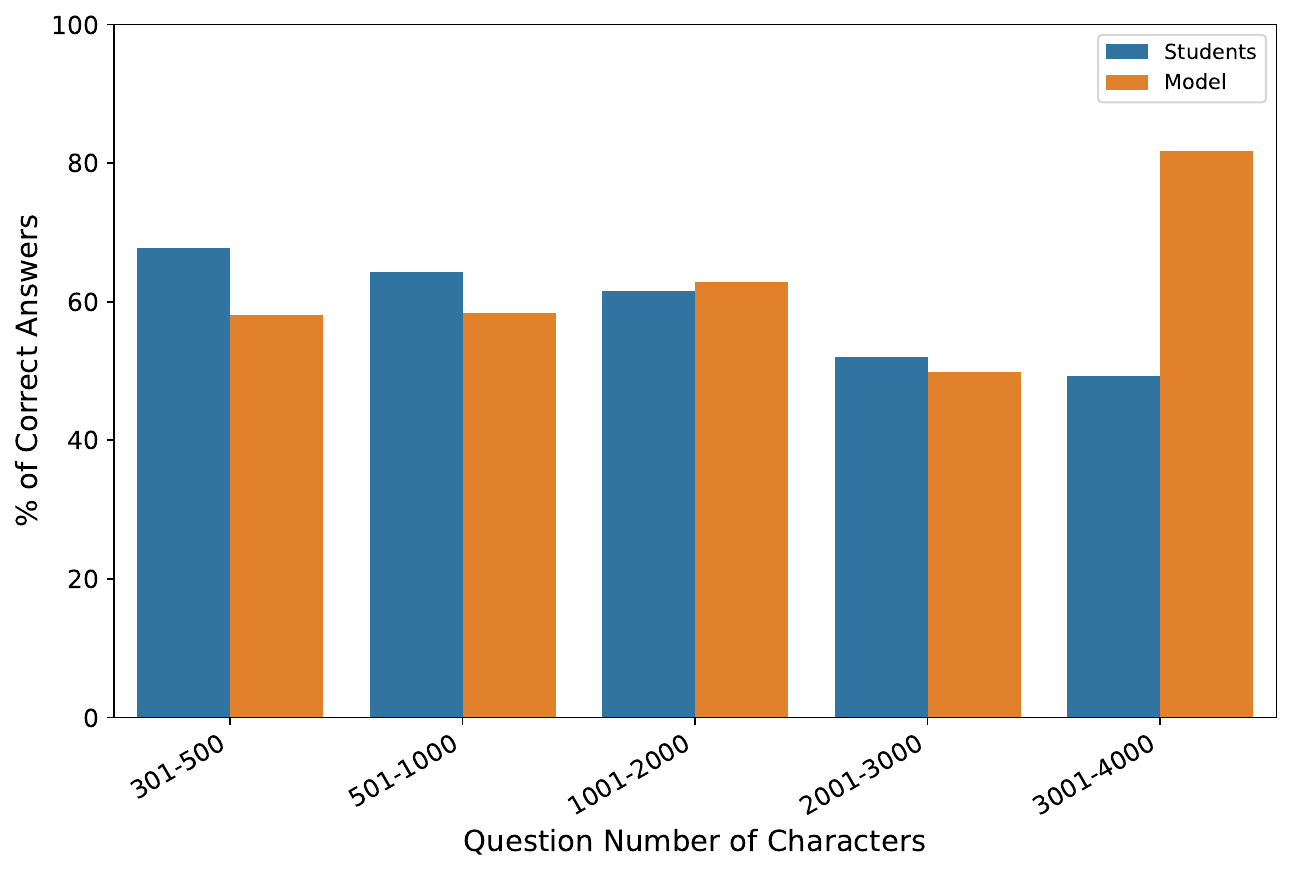}
  \caption{Comparison of student vs model accuracy depending on the question length in characters.}
  \label{fig:char}
\end{figure}

Figure \ref{fig:char} shows the comparison of the student and model accuracy (average majority vote)  depending on the length of the question in characters.

\subsection{Model performance across question and image features}
Model and student performance per question and image features with 95\% confidence intervals are presented in Tables \ref{tab:accuracy_model} and \ref{tab:accuracy_student}.

\begin{table*}[h]
\centering
\begin{tabular}{|l|c|c|c|c|}
\hline
\multirow{2}{*}{Category} & \multirow{2}{*}{Label} & \multicolumn{3}{c|}{Model Accuracy and 95 \% CI} \\ \cline{3-5} 
 &  & Accuracy Mean & Lower Bound & Upper Bound \\ \hline
\multirow{5}{*}{Question feature} & Simple & 0.613 & 0.54 & 0.68 \\
 & Complex & 0.503 & 0.35 & 0.65 \\
 & MCQ & 0.663 & 0.59 & 0.77 \\
 & MCQMA & 0.457 & 0.68 & 0.84 \\
 & Compound & 0.650 & 0.55 & 0.74 \\
 \hline
\multirow{6}{*}{Image feature} & Crucial & 0.561 & 0.49 & 0.63 \\
 & Supplemental & 0.74 & 0.60 & 0.86 \\ 
 & Algorithm & 0.875 & 0.63 & 1.00\\
 & Diagram & 0.566 & 0.48 & 0.65 \\
 & Picture & 0.687 & 0.54 & 0.84 \\
 & Line Plot & 0.556 & 0.43 & 0.69 \\ \hline
\end{tabular} \caption{Model accuracy means and 95\% Confidence Intervals. CI is computed with non-parametric bootstrap using 1000 resamples.}\label{tab:accuracy_model}
\end{table*}

\begin{table*}[t!]
\centering
\begin{tabular}{|c|c|c|c|c|}
\hline
\multirow{2}{*}{Category}&\multirow{2}{*}{Label} & \multicolumn{3}{c|}{Student Accuracy and 95 \% CI} \\ \cline{3-5}  
&& Accuracy Mean & Lower Bound & Upper Bound \\ \hline
\multirow{5}{*}{Question feature} &Simple &  0.628 & 0.59 & 0.66 \\
& Complex &  0.622 & 0.54 & 0.70 \\
& MCQ &  0.689 & 0.64 & 0.74 \\ 
& MCQMA &  0.599 & 0.54 & 0.67 \\ 
&Compound & 0.579 & 0.52 & 0.64 \\ 
\hline
\multirow{6}{*}{Image feature} &Crucial & 0.637 & 0.60 & 0.67 \\ 
& Supplemental &  0.588 & 0.51 & 0.67 \\
&Algorithm &  0.534 & 0.40 & 0.66 \\ 
& Diagram &  0.613 & 0.57 & 0.66 \\ 
& Picture & 0.648 & 0.57 & 0.71 \\ 
& Line Plot & 0.645 & 0.58 & 0.71 \\ 
\hline
\end{tabular}
\caption{Student accuracy means and 95\% Confidence Intervals for different image types. CI is computed with the non-parametric bootstrap using 1000 resamples}
\label{tab:accuracy_student}
\end{table*}

\end{document}